\documentclass[%
 reprint,
bibnotes,
 amsmath,amssymb,
 aps,
]{revtex4-1}

\usepackage{graphicx}% Include figure files
\usepackage[english]{babel}
\usepackage{microtype}          
\usepackage[breaklinks=true,colorlinks=true,linkcolor=blue,urlcolor=blue,citecolor=blue]{hyperref}

\usepackage{xcolor}     %% required only for the \todo command during the development 
\usepackage{mathtools}
\usepackage{braket}
\usepackage{tikz}
 \usepackage{printlen}

 \usepackage{longtable}
  \usepackage{booktabs}
 \usepackage{siunitx}
 \usepackage{csquotes}

\newcommand{\op}[1]{\ensuremath{\hat{#1}}}

\usepackage{dcolumn}% Align table columns on decimal point
\usepackage{bm}% bold math
\begin{document}
\bibliographystyle{apsrev}
\preprint{APS/123-QED}

\title{\textit{Ab initio} Quantum Monte Carlo simulation\\ of the warm dense electron gas in the thermodynamic limit}% Force line breaks with \\

\author{Tobias Dornheim$^{1,\dagger}$}
 \email{dornheim@theo-physik.uni-kiel.de}
\author{Simon Groth$^{1,\dagger}$}%
\author{Travis Sjostrom$^{2}$}
\author{Fionn D.~Malone$^{3}$}
\author{W.M.C.~Foulkes$^{3}$}
\author{Michael Bonitz$^{1}$}
\affiliation{$^\dagger$These authors contributed equally to this work.\\ $^1$Institut f\"ur Theoretische Physik und Astrophysik, Christian-Albrechts-Universit\"{a}t zu Kiel, D-24098 Kiel, Germany\\ $^2$Theoretical Division, Los Alamos National Laboratory, Los Alamos, New Mexico 87545, USA\\ %$^3$Helmholtz-Zentrum Dresden-Rossendorf, D-01328 Dresden, Germany\\
$^{3}$Department of Physics, Imperial College London, Exhibition Road, London SW7 2AZ, UK}

\date{\today}% It is always \today, today,

\begin{abstract}
We perform \emph{ab initio} quantum Monte Carlo (QMC) simulations of the warm dense uniform electron gas in the thermodynamic limit. By combining QMC data with linear response theory 
we are able to remove finite-size errors from the potential energy over the entire warm dense regime, overcoming the deficiencies of the existing finite-size corrections by Brown \emph{et al.}~[PRL \textbf{110}, 146405 (2013)].
Extensive new QMC results for up to $N=1000$ electrons enable us to compute the potential energy $V$ and the exchange-correlation free energy $F_{xc}$ of the macroscopic electron gas with an unprecedented accuracy of $|\Delta V|/|V|, |\Delta F_{xc}|/|F|_{xc} \sim  10^{-3}$. A comparison of our new data to the recent parametrization of $F_{xc}$ by Karasiev {\em et al.} [PRL {\bf 112}, 076403 (2014)] reveals significant deviations to the latter.
%\begin{description}
%\item[PACS: 05.30.Fk, 71.10.Ca]
%\item[71.10.Ca]
%\end{description}
\end{abstract}

\maketitle

The uniform electron gas (UEG), consisting of electrons on a uniform neutralizing background, is one of the most important model systems in physics \cite{loos}.
Besides being a simple model for metals, the UEG has been central to the development of linear response theory and more sophisticated perturbative treatments of solids, the formulation of the concepts of quasiparticles and elementary excitations, and the remarkable successes of density functional theory.

The practical application of ground-state density functional theory in condensed matter physics, chemistry and materials science rests on a reliable parametrization of the exchange-correlation energy of the UEG \cite{perdew}, which in turn is based on accurate quantum Monte Carlo (QMC) simulation data \cite{alder}.
However, the charged quantum matter in astrophysical systems such as planet cores and white dwarf atmospheres \cite{knudson,militzer} is at temperatures way above the ground state, as are inertial confinement fusion targets \cite{nora,schmit,hurricane3}, laser-excited solids \cite{ernst}, and 
%other extreme states of matter produced in laser and ion beam compression experiments 
pressure induced modifications of solids, such as insulator-metal transitions \cite{mazzola14,knudson15}.
This unusual regime, in which strong ionic correlations coexist with electronic quantum effects and partial ionization, has been termed ``warm dense matter'' and is one of the most active frontiers in plasma physics and materials science.

The warm dense regime is characterized by the existence of two comparable length scales and two comparable energy scales.
The length scales are the mean interparticle distance, ${\bar r}$, and the Bohr radius, $a_0$; the energy scales are the thermal energy, $k_B T$, and the electronic Fermi energy, $E_F$.
The dimensionless parameters \cite{wdm} $r_s={\bar r}/a_0$ and $\Theta= k_BT/E_F$ are of order unity.
Because $\Theta \sim 1$, the use of ground-state density functional theory is inappropriate and extensions to finite $T$ are indispensible; these require accurate exchange-correlation functionals for finite temperatures \cite{karasiev2,dharma,gga,burke,burke2}.
Because neither $r_s$ nor $\Theta$ is small, there are no small parameters, and weak-coupling expansions beyond Hartree-Fock
\begin{figure}[] \centering
\hspace*{0.07cm}\includegraphics[width=0.48\textwidth]{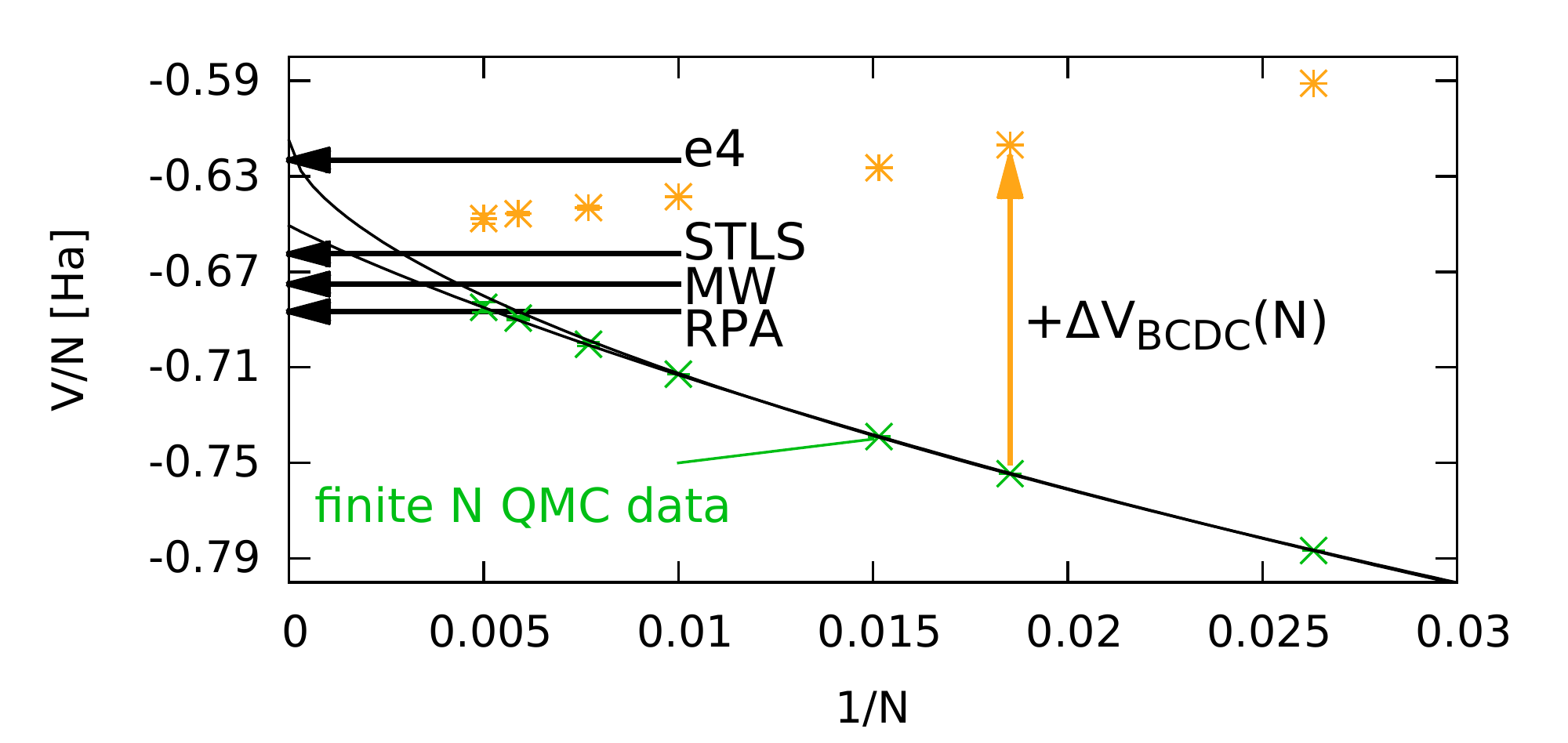}\vspace*{-0.3cm}
\caption{\label{intro_pic}Potential energy per particle of the unpolarized UEG
at $\theta=2$ and $r_s=0.5$.
The exact CPIMC results for different system sizes are indicated by green crosses; the yellow asterisks show these results after the $\Delta V_{\rm BCDC}$ finite-size correction from Eq.~(\ref{EQ_BCDC}) has been applied.
The horizontal arrows show the results of various many-body theories (RPA, STLS \cite{stls}, Montroll-Ward [MW], and $e^4$ [e4] \cite{private}; see text).  The black lines are two different, equally plausbile, extrapolations of the QMC data to infinite system size~\cite{fitnote}. }
\end{figure}
such as the Montroll-Ward (MW) and $e^4$ (e4) approximations~\cite{kremp_springer,vorberger_pre04}, 
as well as linear response theory within the random-phase approximation (RPA) break down \cite{gupta,pdw84}.
Finite-$T$ Singwi-Tosi-Land-Sj\"olander (STLS) \cite{stls,stls2} local-field corrections allow for an extension to moderate coupling \cite{stls2}, but exhibit non-physical behavior 
%(e.g.,~negative values of the pair-correlation function) 
at short distances for moderate to low densities, so improved expressions are highly needed.
Further, quantum-classical mapping \cite{rpa0,dwp} allows for  semi-quantitative descriptions of warm dense matter in limiting cases.

Therefore, an accurate description of warm dense matter in general and of the warm dense UEG in particular can only be achieved using computational approaches, primarily quantum Monte-Carlo (QMC) methods which, however, are hampered by the fermion sign problem \cite{loh,troyer}.
%, which leads to an exponential increase of computation time with system size.
The pioneering QMC simulations of the warm dense UEG by Brown \emph{et al.}\ \cite{brown} eliminated the sign problem by invoking the (uncontrolled) fixed-node approximation~\cite{node}, but were nevertheless restricted to small systems of $N=33$ (spin-polarized) and $N=66$ (unpolarized) electrons and to moderate densities, $r_s \geq 1$.
Recently, we were able to show \cite{groth,dornheim3,malone2} that accurate simulations of these systems are possible over a broad parameter range without any nodal restriction.
Our approach combines two independent methods, configuration path-integral Monte Carlo (CPIMC) \cite{tim1,tim2,prl} and permutation blocking PIMC \cite{dornheim,dornheim2}, which allow for accurate simulations at high ($r_s\lesssim1$) and moderate densities ($r_s\gtrsim1$ and $\theta\gtrsim0.5$), respectively.
An independently developed third approach, density matrix QMC \cite{blunt,malone}, confirmed the excellent quality of these results.
The only significant errors remaining are finite-size effects \cite{fraser,drummond,lin,chiesa,prl,krakauer}, which arise from the difference between the small systems simulated and the infinite [thermodynamic limit (TDL)] system of interest.

Direct extrapolation to the TDL  \cite{alder,drummond,lin} is extremeley costly and also unreliable unless the form of the function to be extrapolated is known; the two black lines in Fig.~\ref{intro_pic} show two equally reasonable extrapolations \cite{fitnote} that reach different limits.
Furthermore, the parameter-free finite-size correction (FSC) proposed in Ref.~\cite{brown} (see Eq.~(\ref{EQ_BCDC}) below) turns out to be inappropriate in parts of the warm dense regime.
The problem is clear from inspection of the yellow asterisks in Fig.~\ref{intro_pic}, which include this FSC but remain system-size dependent.

In this letter, we close the gap between the finite-$N$ QMC data and the TDL by deriving a highly accurate FSC for the interaction energy.
This allows us to obtain precise (on the level of $0.1\%$) results for the  exchange-correlation free energy, making possible the \emph{ab initio} computation of arbitrary thermodynamic quantities over the entire warm dense regime.

\begin{figure}[]
 \centering
\includegraphics[width=0.48\textwidth]{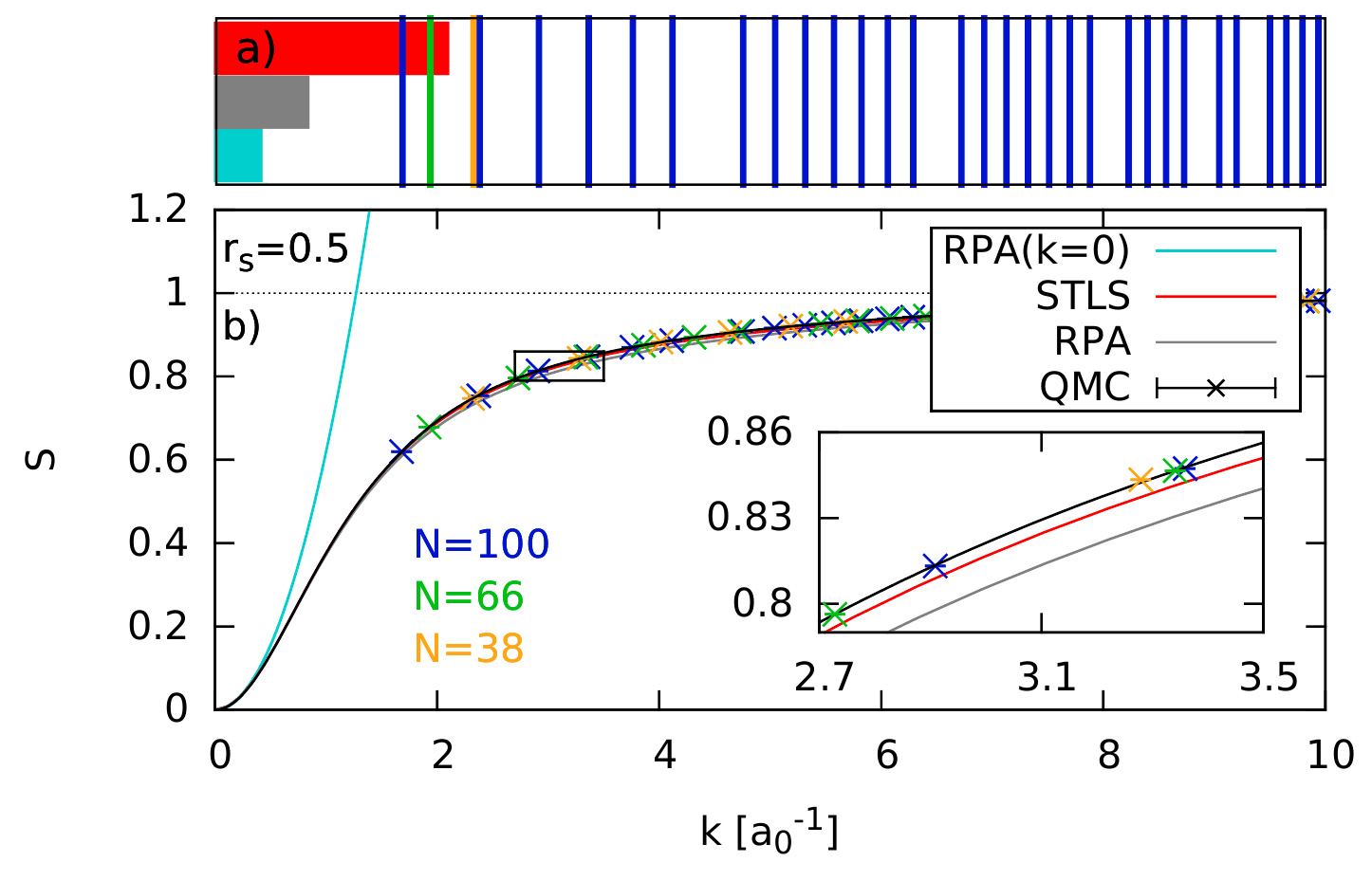}
 \caption{\label{main_Sk}Static structure factors for $\theta=2$, $r_s = 0.5$ and three values of $N$. 
In panel a), the discrete QMC $k$-points are plotted as vertical lines for $N=100$, the minimum k-values for $N=66$ and $N=38$ are indicated by the green and yellow line. The colored horizontal bars indicate the $k$-ranges where $S^{\rm STLS}$ (red), $S^{\rm RPA}$ RPA (grey) and $S^{\rm RPA}_0$ (light blue) are accurate.
Panel b) shows that the QMC results for $S(k)$ converge rapidly with $N$ (see the colored symbols in the inset).
The black curve shows $S_{\rm comb}$ connecting $S^{\rm STLS}(k)$ at small $k$ with the QMC data for $N=100$ which yields accurate results for all $k$.
}
\end{figure}

{\bf Theory}. Consider a finite unpolarized UEG of $N$ electrons subject to periodic boundary conditions.
The Hamiltonian is $\op{H}=\op{K}+\op{V}_\textnormal{E}$, where $\op{K}$ is the kinetic energy of the $N$ electrons in the cell and 
\begin{equation}
\hat{V}_\textnormal{E} = \frac{1}{2}\sum_{i\ne k}^N 
\phi_\textnormal{E}(\mathbf{r}_i, \mathbf{r}_k) 
+ \frac{1}{2}N\xi_\textnormal{M}
\end{equation}
is the Coulomb interaction energy per unit cell of an infinite periodic array of images of that cell.
The Ewald pair potential $\phi_\textnormal{E}(\mathbf{x},\mathbf{y})$ and Madelung constant $\xi_\textnormal{M}$ are defined in Refs.~\cite{fraser,drummond}.
We use Hartree atomic units throughout this work.
The expected value of $\hat{V}_{\textnormal{E}}/N$ carries a finite-size error \cite{tak_note} that is the difference between the potential energy $v$ per electron in the infinite system and its value $V_N/N$ in the finite system. This difference may be expressed in terms of the static structure factor (SF) as follows:
\begin{align}
&\frac{\Delta V_N[S(k), S_N(\mathbf{G})]}{N} =  \underbrace{ \frac{1}{2}\int_{k<\infty}\frac{\textnormal{d}\mathbf{k}}{(2\pi)^3} \left[S(k)-1\right]\frac{4\pi}{k^2} }_{v} \notag \\  \label{delta_V_1}
& \hspace*{5em} - \underbrace{ \left(
 \frac{1}{2L^3}\sum_{\mathbf{G}\ne\mathbf{0}}\left[S_N(\mathbf{G})-1\right]\frac{4\pi}{G^2}+\xi_\textnormal{M}\right) }_{V_N/N} , 
\end{align}
where $L$ and $\mathbf{G}$ are, respectively, the length and reciprocal lattice vector of the simulation cell, $S(k)$ [$S_N(\mathbf{G})$] is the SF of the infinite [finite] system.
A first source of FS error in Eq.~(\ref{delta_V_1}) is the replacement of $S(k)$ in the first term by its finite-size analogue $S_N(\mathbf{G})$ in the second term. However, this effect is negligible, as we will demonstrate in Fig.~\ref{main_Sk}.

Thus the main source of error is the discretization of the integral in the first term to obtain the sum in the second term.
In fact, Chiesa \textit{et al.}~\cite{chiesa}~suggested that the main contribution to Eq.~(\ref{delta_V_1}) comes from the omission of the $\mathbf{G}=\mathbf{0}$ term from the summation~\cite{madelung}.
As is well known, the RPA becomes exact in the limit of small $k$, and the expansion of $S(k)$ around $k=0$ at finite $T$ is given by~\cite{rpa0}
\begin{equation}
\label{EQ_RPA_zero}S^{\rm RPA}_0(k) = \frac{k^2}{2\omega_p}\textnormal{coth}\left( \frac{\beta\omega_p}{2} \right) ,
\end{equation}
where $\beta=1/k_\textnormal{B}T$ is the inverse temperature and $\omega_p=\sqrt{3/r_s^3}$ is the plasma frequency.
The finite-$T$ version \cite{fs_zero} of the Chiesa FSC~\cite{brown},
\begin{equation}
\label{EQ_BCDC}\Delta V_\textnormal{BCDC}(N) = \lim_{k\rightarrow 0} 
\frac{S_0^{RPA}(k)4 \pi}{2 L^3 k^2} = \frac{\omega_p}{4N} \textnormal{coth}\left( \frac{\beta\omega_p}{2} \right) ,
\end{equation}
would be sufficient if: (i) $S^{\rm RPA}_0(k)$ were accurate for the smallest nonzero $k$ in the QMC simulation, $k_\textnormal{min}=2\pi/L$; and (ii) all contributions to Eq.~(\ref{delta_V_1}) not accounted for by the inclusion of the $\mathbf{G} = \mathbf{0}$ term were negligible.
As we demonstrate below, for high temperatures and intermediate to high densities, both conditions are strongly violated.
Thus, we need to use an improved model SF, $S_\textnormal{model}(k)$, to compute the discretization error,
\begin{equation}
  \Delta_N[S_\textnormal{model}(k)] = \frac{\Delta V_N \left[S_\textnormal{model}(k),S_\textnormal{model}(k)\right]}{N} ,\label{EQ_trial}
\end{equation}
in Eq.~(\ref{delta_V_1}).
A natural strategy is to combine the QMC data for $k\geq k_\textnormal{min}$ with an approximation that is accurate for all $k$ up to (at least) $k_\textnormal{min}$.

{\bf Results}. In Fig.~\ref{main_Sk}, we analyze the static SF for $\theta=2$ and a comparatively high density case, $r_s=0.5$, for three different particle numbers.
The use of a finite simulation cell subject to periodic boundary conditions discretizes the momentum, so QMC data are available only at the discrete $k$-points indicated by the vertical lines in the top panel.
As shown in the inset, the QMC $S(k)$ is well converged with respect to system size for surprisingly small $N$, providing justification to set $S_N(\mathbf{G})\approx S(G)$.
Therefore, the FS error of $V_N/N$ reduces as $N$ increases primarily because the $k$-grid becomes finer and $k_\textnormal{min}$ decreases.
The figure also allows us to study the performance of the three analytical structure factors, $S^{\rm RPA}$, $S^{\rm STLS}$ \cite{stls,stls2} and $S^{\rm RPA}_0$.
We clearly observe that $S^{\rm RPA}_0(k)$ is only accurate for $ka_0\lesssim 0.3$, explaining why the BCDC FSC, Eq.~(\ref{EQ_BCDC}), fails.
In contrast, $S^{\rm RPA}(k)$ and  $S^{\rm STLS}(k)$ match the QMC data much better.
%quite well over the entire $k$-range.
On the left-hand side of panel a), we indicate the $k$-ranges over which the three models are accurate, 
showing that only $S^{\rm STLS}(k)$ connects smoothly to the QMC data. At larger $k$, $S^{\rm RPA}$ and $S^{\rm STLS}$ exhibit significant deviations from the QMC data, although STLS is more accurate. For completeness, we mention that, when the density is lowered, the $k-$ranges of accurate behavior of 
$S^{\rm RPA}$, $S^{\rm STLS}$ and $S^{\rm RPA}_0$ continuously increase \cite{supplement}. For example, at $r_s=1$, both $S^{\rm RPA}$ and $S^{\rm STLS}$ smoothly connect to the QMC data whereas for $r_s=10$ this is observed even for $S^{\rm RPA}_0(k)$ revealing that there the BCDC FSC is accurate.

Based on this behavior, an obvious way to construct a model SF that is accurate over the entire $k$-range for all warm dense matter parameters is to combine the QMC data with the STLS data at small $k$. The result is denoted $S_\textnormal{comb}$ and computed via a spline function. The excellent behavior is illustrated by the black line in panel b) of Fig.~\ref{main_Sk} and in the inset.
This quasi-exact SF is the proper input to compute the discretization error from Eq.~(\ref{EQ_trial}).
 
\begin{figure}[]
 \centering
\includegraphics[width=0.45\textwidth]{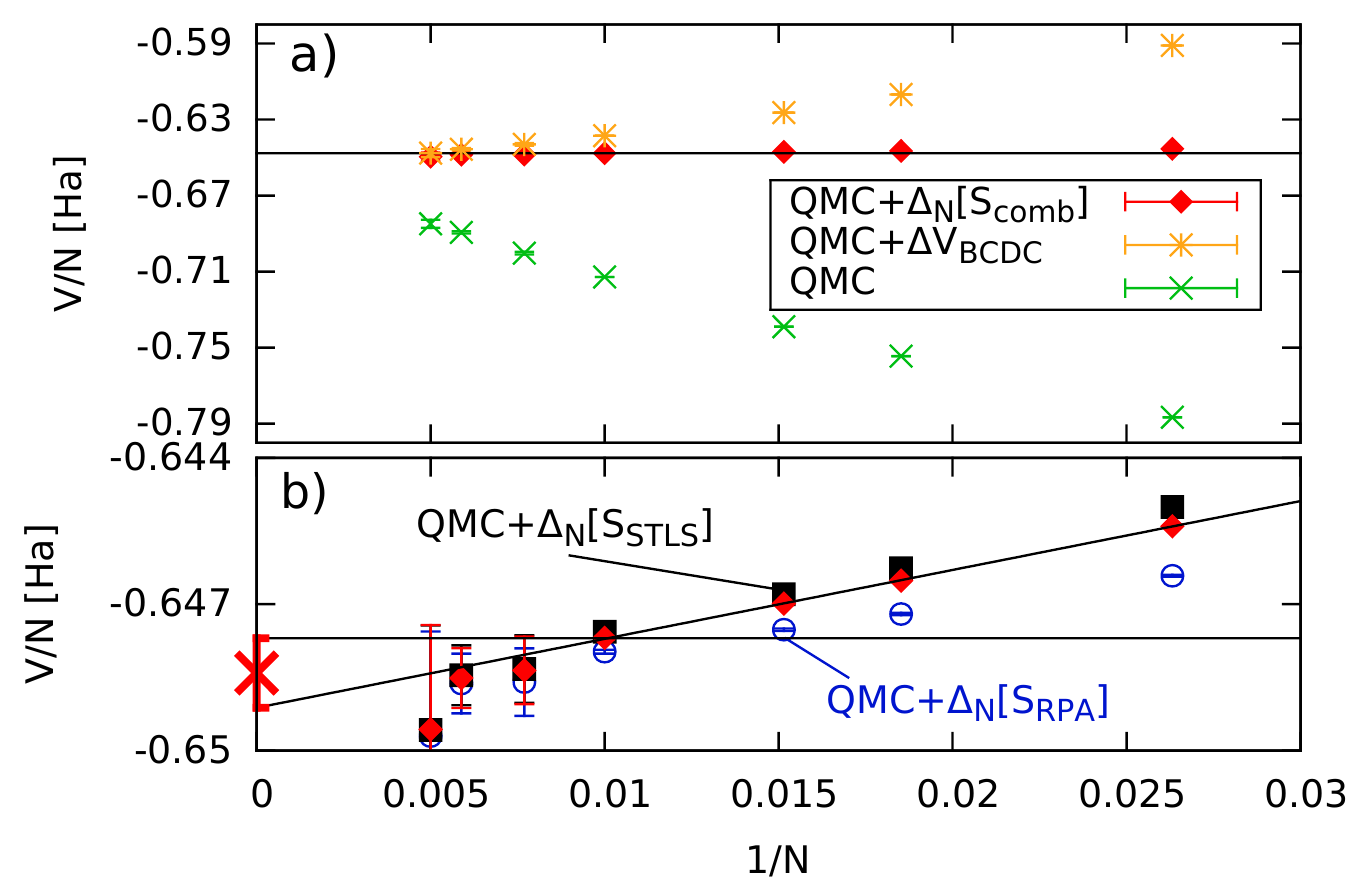}
 \caption{\label{rs05} {\bf a)} Finite-size corrected QMC data for the potential energy for $\theta=2$ and $r_s=0.5$. The yellow asterisks are obtained using Eq.~(\ref{EQ_BCDC}); the red diamonds  use the combined SF $S_{\textnormal{comb}}$ (cf. Fig.~\ref{main_Sk}) to evaluate the discretization error, Eq.~(\ref{EQ_trial}).
{\bf b)} Magnified part of panel a) including an extrapolation of the residual finite-size error to the TDL (the red cross).
Results obtained using only the full RPA (blue) and STLS structure factors (black) in Eq.~(\ref{EQ_trial}) are also shown.
}
\end{figure}
 
The results of this procedure are shown in Fig.~\ref{rs05} for the most challenging high-density case, $r_s=0.5$ and $\theta=2$.
Clearly, the raw QMC data (green crosses) suffer from severe finite-size errors of order $10\%$ for system sizes from $N=38$ to $N=200$. These errors do not exhibit the $\Delta V \propto 1/N$ behavior predicted by Eq.~(\ref{EQ_BCDC}), and the BCDC-corrected QMC data (yellow asterisks) do not fall on a horizontal line.
In contrast, using $\Delta_N[S_{\textnormal{comb}}]$ 
produces results that are very well converged for all system sizes considered, including even $N=38$ (red diamonds).
Panel b) of Fig.~\ref{rs05}  shows that the removal of the discretization error has reduced the FS bias by two orders of magnitude.
The residual error, $|\Delta V|/|V|\sim10^{-3}$, is due to the small finite-size effects in the QMC data for $S_N(k)$ itself and exhibits a linear behavior in $1/N$.
Thus, it is possible to determine the potential energy in the TDL (the red cross in the bottom panel) with a reliable error bar \cite{extra_note}.

To further explore the properties of our discretization formula for the FS error, we recompute $\Delta_N$ using the purely theoretical STLS and RPA SFs as $S_{\textnormal{model}}$ in Eq.~(\ref{EQ_trial}). 
The FS-corrected data are depicted by the black squares and blue circles in panel b) of Fig.~\ref{rs05}, respectively.
Surprisingly, we find very good agreement with the FSCs derived from the substantially more accurate $S_\textnormal{comb}$.
Hence, despite their significant deviations from the QMC data at intermediate $k$ (cf.~inset in panel b of Fig.~\ref{main_Sk}), $S^{\rm STLS}(k)$ and $S^{\rm RPA}(k)$ are sufficiently accurate to account for the discretization error of the potential energy \cite{explanation}.
Since $S_\textnormal{comb}$ is sensitive to statistical noise, computing the FSC solely from $S^{\rm STLS}(k)$ or $S^{\rm RPA}(k)$ is in fact the preferred approach.
Of course, this unexpectedly simple solution to the finite-size-correction problem does not eliminate the need for accurate finite-$N$ QMC data, the quality of which sets the base line for our thermodynamic result, $v=V_{\textnormal{QMC},N}/N +\Delta_N[S_\textnormal{model}]$.
Using instead the STLS or RPA SF to estimate $V_{\textnormal{QMC},N}$ as well as $\Delta_N$ poorly accounts for  the short-range correlations and, even for $\theta=2$ and $r_s=0.5$, leads to $\sim 10\%$ errors (cf.~Fig.~\ref{intro_pic}), which further increase with $r_s$.
\begin{figure}[]
 \centering
\includegraphics[width=0.44\textwidth]{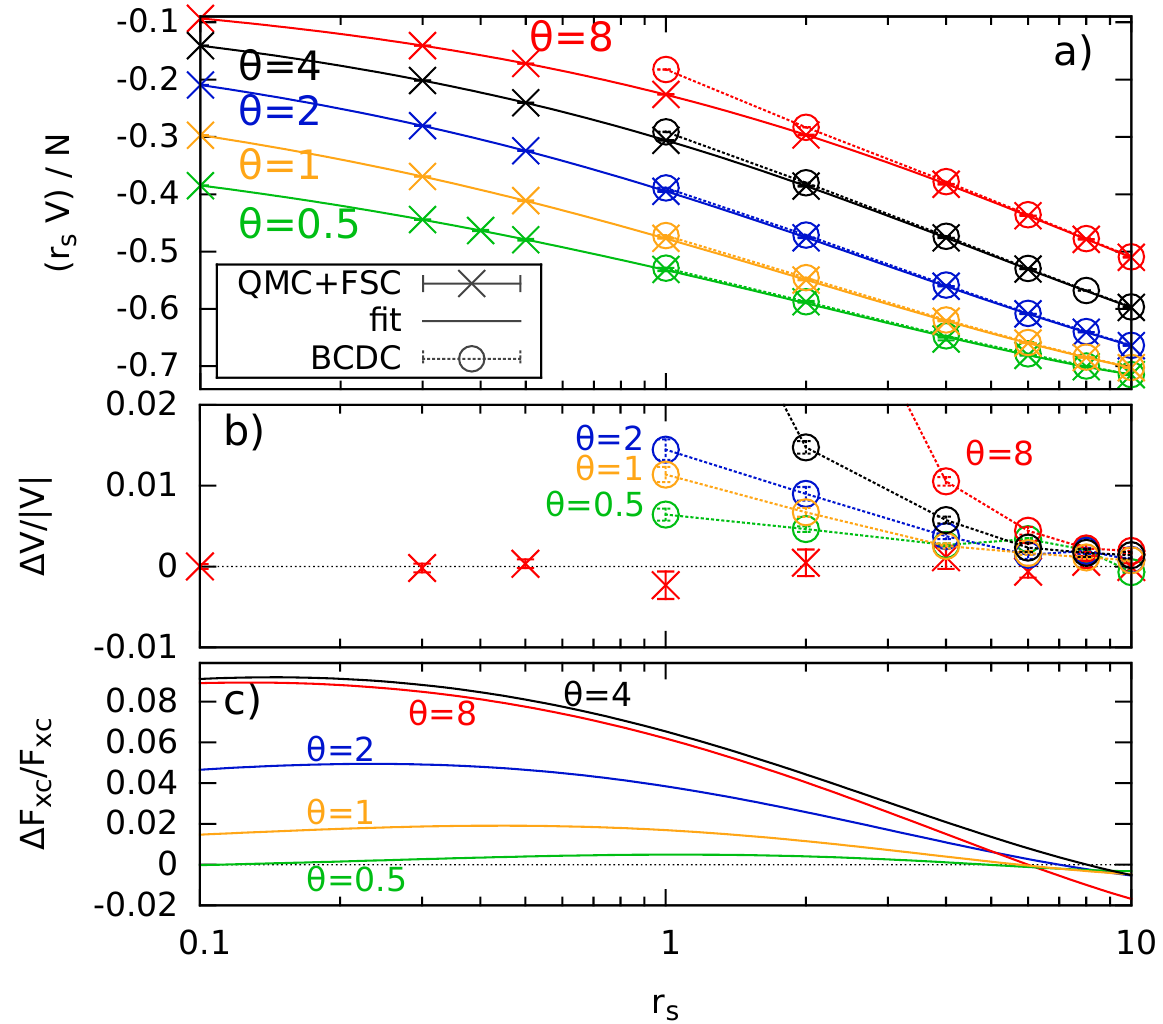}\vspace*{-0.22cm}
 \caption{\label{Pot} 
Potential energy of the UEG in the TDL.
\textbf{a)}~Our new FS-corrected QMC data, the fits to our data (see Eq.~(S.2) of Ref.~\cite{supplement}), and the RPIMC results of Brown \emph{et al.} \cite{brown_error}, which include BCDC FSCs.
\textbf{b)}~Relative deviations of our data (for $\Theta = 8$) and Brown's BCDC-corrected data from the corresponding fit. {\bf c)} Relative deviation of our exchange correlation free energies from the fit of Ref.~\cite{karasiev} for five temperatures. For details see Ref.~\cite{supplement}.
}
\end{figure}

By performing extensive QMC simulations and applying our FSC to results for various system sizes $N$ to allow extrapolatation of the residual FS error, we obtain the potential energy of the UEG in the TDL over a very broad density range, $0.1 \le  r_s \le 10$.
The results are displayed in Fig.~\ref{Pot} for five different temperatures and listed in a table in the supplement \cite{supplement}.
We also compare our results to the most accurate data previously available --- the  RPIMC results of Brown \textit{et al.}~(BCDC, circles), which were corrected using the BCDC FSC, Eq.~(\ref{EQ_BCDC}) \cite{brown_error}.
We underline that these results were limited to moderate densities, $r_s\ge1$, but even there substantially deviate from our data.
The error increases rapidly with density and temperature reaching $20\%$ for $r_s=1$ and $\theta=8$ \cite{supplement}.

Finally, we obtain the exchange-correlation free energy from a fit to the potential energy, regarded as a function of $r_s$ for fixed $\theta$.
Panel b) of Fig.~\ref{Pot} shows that the functional form assumed (Eq.~(S.2) in Ref.~\cite{supplement}) is indeed appropriate as no systematic deviations between the QMC data and the fit (red crosses, $\theta=8$) are observed.
In panel c) of Fig.~\ref{Pot}, we compare our new data for $F_{xc}$ to the recent parametrization by Karasiev {\em et al.}
~\cite{karasiev}.
By design, both curves coincide in the limit $r_s\to0$, approaching the exact asymptotic value known from Hartree-Fock theory (for $r_s \ll 0.1$).
While both results are in very good agreement for $\theta=0.5$, we observe severe deviations of up to  $9\%$ at $\theta=8$ [$5\%$ at $\theta=2$].
Despite the systematic RPIMC bias and the lack of data for $r_s<1$ prior to our work, the major cause of the disagreement is the inadequacy of the BCDC FSCs for high temperature and small $r_s$.
The absolute data for $F_{xc}$ and the corresponding fit parameters are provided in Ref.~\cite{supplement}.

{\bf Summary and discussion}.
We have presented a simple but highly accurate procedure for removing finite-size errors from {\em ab initio} finite-$N$ QMC data for the potential energy $V$ of the UEG at finite temperature.
This is achieved by adding to the QMC results the discretization error $\Delta_N[S_\textnormal{model}(k)]$, Eq.~(\ref{EQ_trial}), computed using simple approximate structure factors based on the RPA or STLS approximations.
Our finite-size-corrected results include excellent descriptions of both the exchange and short-range correlation effects (from the QMC data) and the long-range correlations (via the RPA or STLS corrections).
These results constitute the first {\em ab initio} thermodynamic data for the warm dense electron gas free of the limitations of many-body approximations or systematic simulation biases such as fixed-node error.
For temperatures above half the Fermi temperature and a density range covering six orders of magnitude ($0.1 \le r_s \le 10$), we achieve an unprecedented accuracy of $\sim 0.3\%$; our results will therefore serve as valuable benchmarks for the development of accurate new theories and simulation schemes, including improved static local field corrections.
In particular, we observe that the recent results of Brown \textit{et al.}~\cite{brown,brown_error}, which were obtained by applying the BCDC FSC from Eq.~(\ref{EQ_BCDC}) to RPIMC data, exhibit deviations of up to $20\%$.
The recent parametrization of $F_{xc}$ by Karasiev \textit{et al.}~\cite{karasiev}, which was mainly based on the data by Brown \textit{et al.}, uses a good functional form but exhibits errors of up to $9\%$ at high temperatures.
Even though these inaccuracies constitute only a small fraction of the total free energy, which might not drastically influence subsequent DFT calculations of realistic multi-component systems, it is indispensable to have a reliable and consistent fit of $F_{xc}$ for all WDM parameters to achieve predictive power and  agreement with experiments.
The construction of an improved complete parametrization of $F_{xc}$ with respect to density, temperature, and spin polarization remains a challenging task for future work. 
In particular, the fermion sign problem presently limits our QMC simulations to $\theta\geq0.5$ for $r_s\sim1$ (although lower temperatures are feasible both for larger and smaller $r_s$ with PB-PIMC and CPIMC, respectively). To overcome this bottleneck, it will be advantageous to incorporate the $T=0$ limit of $E_{xc}$ and, thus, to perform an interpolation across the remaining gap where no ab initio data are available.
In addition to finite-$T$ DFT, we expect such a fit to be of key importance as input for quantum hydrodynamics \cite{manfredi, michta} and time-dependent DFT.
Finally, our FSC procedure is expected to be of value for other simulations of warm dense plasmas \cite{driver1,driver2,driver3}, as well as $2D$ systems, e.g.~Refs.~\cite{2d_1,2d_2}.

\section*{Acknowledgements}
We acknowledge stimulating discussions with Tim Schoof and are grateful to Jan Vorberger for providing the Montroll-Ward and e$^4$ data shown in Fig.~\ref{intro_pic}.
This work was supported by the Deutsche Forschungsgemeinschaft via project BO1366-10 and via SFB TR-24 project A9 as well as grant shp00015 for CPU time at the Norddeutscher Verbund f\"ur Hoch- und H\"ochstleistungsrechnen (HLRN).
TS~acknowledges the support of the US DOE/NNSA under Contract No.~DE-AC52-06NA25396.
FDM is funded by an Imperial College PhD Scholarship.
FDM and WMCF used computing facilities provided by the High Performance Computing Service of Imperial College London, by the Swiss National Supercomputing Centre (CSCS) under project ID s523, and by ARCHER, the UK National Supercomputing Service, under EPSRC grant EP/K038141/1 and via a RAP award.
FDM and WMCF acknowledge the research environment provided by the Thomas Young Centre under Grant No.~TYC-101.


\begin{thebibliography}{26}
\bibitem{wdm} The third condition of strong ionic coupling is not  relevant for our analysis of the UEG.
\bibitem{fitnote} The two different functional forms for the fits in Fig.~\ref{rs05} are $f(1/N) = a_f + {b_f}/{N^{c_f}}$ and $g(1/N) = a_g + {b_g}/{N} + {b_c}/{N^{b_d}}$.
\bibitem{tak_note}  In the considered temperature range shell-filling effects are negligible and twist-averaging~\cite{lin,drummond} is not required. Note that $V$ is sufficient to obtain $F_{xc}$ which gives access to all other quantities, including the kinetic energy.
\bibitem{madelung} The Madelung constant is approximated by \cite{drummond}
\begin{equation}
\quad\xi_\textnormal{M}\approx \frac{1}{L^3} \sum_{\mathbf{G}\neq \mathbf{0}} \frac{4\pi}{G^2} e^{-\epsilon G^2}
- \frac{1}{(2\pi)^3}\int_{k < \infty}\textnormal{d}\mathbf{k}\frac{4\pi}{k^2}e^{-\epsilon k^2}
\nonumber
\end{equation}
for small $\epsilon$,
and, therefore, cancels the minus unity contributions to both the sum and the integral in Eq.~(\ref{delta_V_1}).
\bibitem{fs_zero} Note that in the limit $T\to0$ ($\beta\to\infty$) the hyperbolic cotangent function in EQ.~(\ref{EQ_BCDC}) approaches unity and thus the ground state result from Refs.~\cite{chiesa,drummond} is recovered.
\bibitem{extra_note}
We (i) perform a linear extrapolation and (ii) average over the last few data points where $V/N$ are indistinguishable within the statistical error bars. Our final result for $V/N$ is the mean of (i) and (ii), which constitute reasonable lower and upper bounds, respectively.
\bibitem{explanation}
The discretization FSC is insensitive to the choice of SF because the discretization error is unaffected by constant shifts in the model SF (the three SFs in the inset of Fig.~\ref{main_Sk} are nearly parallel).
\end{thebibliography}
\end{document}